\documentstyle[twoside,fleqn,espcrc2,epsf]{article}
\newcommand{\be}{\begin{equation}}
\newcommand{\ee}{\end{equation}}
\newcommand{\bea}{\begin{eqnarray}}
\newcommand{\eea}{\end{eqnarray}}
\newcommand{\bean}{\begin{eqnarray*}}
\newcommand{\eean}{\end{eqnarray*}}

\newcommand{\AmS}{{\protect\the\textfont2
  A\kern-.1667em\lower.5ex\hbox{M}\kern-.125emS}}

\title{Relativistic Quarkonia from Anisotropic Lattices
\thanks{Special thanks to R.~Mawhinney for presenting this talk 
in a last-minute change due to exceptional circumstances.
This work was conducted on the QCDSP machines at 
Columbia University and RIKEN-BNL Research Center. 
TM and XL are supported by DOE. 
}}

\author{ 
P.~Chen\address{Physics Department, Columbia University, 
New York, NY 10027, USA}, X.~Liao$^{\rm a}$, 
T.~Manke$^{\rm a}$ }
\begin{document}

\begin{abstract}
We report on new results for the spectrum of quarkonia 
using a fully relativistic approach on anisotropic lattices 
with quark masses in the range from strange to bottom.
A fine temporal discretisation also enables us to resolve excitations 
high above the ground state. In particular we studied the mass dependence 
and scaling of hybrid states.
\end{abstract}

%
\maketitle
\section{INTRODUCTION}

The study of heavy quark systems on the lattice is complicated
by the large separation of momentum scales which are difficult to
accommodate on isotropic lattices.
Over the years many effective descriptions of QCD
have been developed for low energies and tested against 
a wealth of experimental data \cite{nrqcd,nrqcd_omv4,fermilab}. 
However, the \mbox{(non-)}perturbative control of higher dimensional operators 
is very difficult and in practise one has to rely on 
additional approximations. Even in non-relativistic bottomonium calculations
one has observed sizable relativistic corrections to the spin structure
\cite{nrqcd_omv6,nrqcd_spitz}
and, more worrying, large scaling violations \cite{nrqcd_nf2} 
which cannot be controlled
by taking the continuum limit.
We take this as our motivation to study heavy quarkonia on anisotropic lattices in a fully relativistic framework.

The strategy and first results for charmonium have already been presented in 
\cite{tim,ping}. The basic idea is to control large lattice spacing artefacts from the heavy quark mass by adjusting the temporal lattice spacing, $a_t$, 
so that $m_q a_t < 1$.
The continuum limit can then be taken at fixed anisotropy, $\xi = a_s/a_t$.
The details of our calculation are given in Section 2.

In our study we also employ another advantage of anisotropic lattices -
a fine temporal resolution is ideal to trace the fast exponential fall-off
from higher excited states. In particular the gluonic excitations have
attracted considerable interest as they give rise to exotic quantum numbers not allowed in the quark model. We present our results for such exotic hybrid
states in Section 3.
Section 4 concludes this report with results from a first 
relativistic  bottomonium calculation.

\section{ANISOTROPIC LATTICES}
To study the relativistic propagation of heavy quarks it is mandatory
to have a fine resolution in the temporal lattice direction.
To this end we employ an anisotropic gluon action:

\bea
S = - \beta \left( \sum_{x, {\rm i > j}} \xi_0^{-1} P_{\rm i j}(x) +
    \sum_{x, {\rm i}}     \xi_0      P_{\rm i t}(x) \right) ~~.
\label{eq:aniso_glue}
\eea

This is the standard Wilson action written in terms of simple plaquettes,
$P_{\mu\nu}(x)$. Here $(\beta,\xi_0)$ are two bare parameters, which 
determine the gauge coupling and the renormalised anisotropy, 
$\xi = a_s/a_t$, of the quenched lattice. 
The anisotropic gluon action (\ref{eq:aniso_glue}) is designed to be accurate 
up to ${\cal O}(a_s^2,a_t^2)$. 

For the heavy quark propagation in the gluon background 
we used the ``anisotropic clover'' formulation as first described in 
\cite{tim,ping}. The discretised form of the continuum Dirac operator, 
$Q=m_q+D\hskip -0.21cm \slash $, reads
\bea
Q & = & m_0 + \nu_s~W_i \gamma_i + \nu_t~W_0 \gamma_0 - \nonumber \\ 
& & \frac{a_s}{2}\left[ c_s~\sigma_{0k}F_{0k} + c_t~\sigma_{kl}F_{kl} \right]~~, 
\nonumber \\
W_\mu & = & \nabla_\mu - (a_\mu/2) \gamma_\mu \Delta_\mu ~~.
\label{eq:aniso_quark}
\eea
This is indeed the most general anisotropic quark action 
including all operators to dimension 5 (up to field redefinition).
The five parameters in Eq. \ref{eq:aniso_quark} are all related
to the quark mass, $m_q$, and the gauge coupling as they appear in the
continuum action. Their classical
estimates have been given in \cite{ping}.
Here we chose $m_0$ non-perturbatively, such that the rest energy
of the hadron corresponds to its experimental value
(e.g. \mbox{$M(^{3}S_1^{--})$ =} \mbox{3.097 GeV} for charmonium). We also fix $\nu_s=1$
and adjust $\nu_t$ non-perturbatively for the mesons to obey a relativistic dispersion relation (~$c({\bf 0})=1$~):
\bea
E^2({\bf p}) &=& E^2({\bf 0}) + c^2({\bf p})~{\bf p}^2 + {\cal O} ({\bf p}^4) \ldots ~~.
\eea

For the clover coefficients $(c_s,c_t)$ we choose their classical estimates
and augment this prescription by tadpole improvement. The tadpole coefficient
has been determined from the average link in Landau gauge: $u_{0L} = 1/3~\langle U_\mu(x) \rangle_{\rm Landau}$. Any other choice will give the same
continuum limit, but with this prescription we expect only small
${\cal O}(\alpha a)$ discretisation errors. 

From the quark propagators we construct meson correlators for bound
states with quantum numbers $S(0,1) \times  L(0,1,2)$ and, for 
hybrid states, with additional gluonic excitation.
Our fundamental bilinears are constructed from two 4-spinors, one of 
16 $\Gamma$- matrices and additional insertions of up to 2 lattice 
derivatives:  
\be
M(x) = \bar q(x) \Gamma_i \Delta_j \Delta_k q(x)~~.
\ee

From those basic operators we can construct a vast number of meson states
with different and definite $J^{PC}$. For example, 
the exotic quantum numbers $1^{-+}$ can be obtained from  
$\bar q(x)~\epsilon_{ijk}~ \gamma_j ~\epsilon_{klm} \Delta_l \Delta_m~q(x)$.
These simple-minded operators can be further improved to give states with
different projection onto the ground state. 
Here we follow \cite{nrqcd_hybrid} and employ a combination
of various quark smearings and APE-smearing for the gauge links. 
This allows us to extract reliably both the ground state energies and 
their excitations from correlated multi-state fits to several channels.
We have also checked our parameter estimates for consistent fit results
from different methods and ranges. In order to call a fit acceptable 
we require its Q-value to be bigger than 0.1.
In Table \ref{tab:para} we list the parameters of our simulation.
\begin{table}[b]
\vskip -7mm
\caption{Simulation Parameters. Using $r_0$ to set the scale, we adjust 
the bare quark mass such that the $1^{--}$ corresponds to the 
experimental values in 
$s\bar s (\phi), c \bar c (J/\Psi)$ and  $b \bar b (\Upsilon)$.}
\begin{tabular}{cccc}
\hline
$(\beta,\xi)$ & $N_s^3 \times N_t$ & configs    & $1^{--}$ [GeV]  \\
\hline
(5.7, 2)      & $8^3 \times 32$    & 817        & 1.0180(46)   \\
(5.7, 2)      & $8^3 \times 32$    & 1950       & 3.099(2)   \\
(5.9, 2)      & $16^3 \times 64$   & 1080       & 3.090(1)  \\
(6.1, 2)      & $16^3 \times 64$   & 1010       & 3.062(2)  \\
(6.1, 4)      & $8^3 \times 96$    & 200        & 9.4659(50)  \\
\hline
\end{tabular}
\label{tab:para}
\end{table}

\begin{figure}[t]
\hbox{\epsfxsize = 70mm  \epsfysize = 55mm \hskip -1mm \epsffile{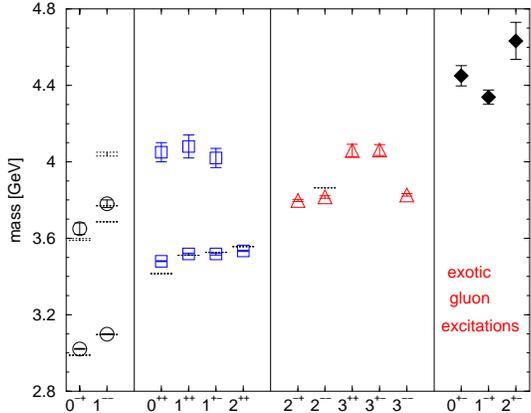}}
\vskip -8mm
\caption{Quenched charmonium spectrum at $(\beta,\xi) = (5.7, 2)$. The lattice spacing is set from $r_0$. Horizontal lines show the experimental values.}
\vskip -9mm
\label{fig:charm_spect}
\end{figure}

A representative compilation of our results for the
charmonium spectrum is shown in Figure \ref{fig:charm_spect}
for one lattice spacing. Most noticeable is the clear level ordering which
can be seen for states with different orbital/gluon angular momentum.
To convert our lattice results into dimensionful quantities we
used the Sommer scale, $r_0$, as defined in \cite{sommer} and calculated
very accurately in \cite{klassen_r0} for isotropic and anisotropic lattices.
It is well-known that, without dynamical sea quarks in the gluon background, 
the definition of the lattice spacing is ambiguous and one cannot 
reproduce all experimental splittings simultaneously.
We are however strongly encouraged by the overall agreement of this 
first-principle calculation with experimental data (where available). 
This bodes well for the reliability of our quenched predictions 
for states which have yet to be seen by experimentalists.
Here we accept the shortcomings of the quenched 
approximation and focus on the scaling behaviour instead.

A detailed study of the spin structure has been presented previously
\cite{tim,ping}. In the following we will analyse the hybrid
excitations in more detail.

\section{HYBRID EXCITATIONS}

The exotic hybrid states in Figure \ref{fig:charm_spect} are of particular 
interest as they reveal explicitly the gluons in the low energy 
regime of QCD. Here we analyse the robustness of our results as we change
the lattice spacing and quark mass in our calculation.

\begin{figure}[t]
\hbox{\epsfxsize = 70mm  \epsfysize = 55mm \hskip -1mm \epsffile{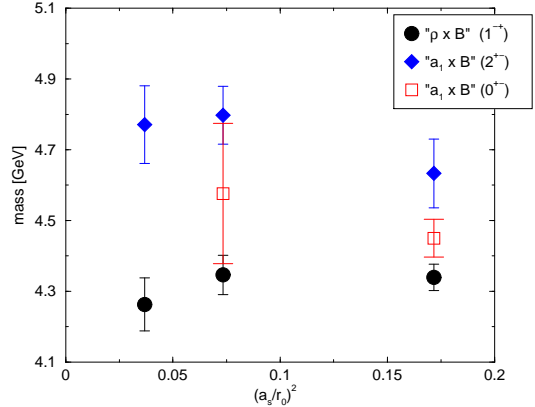}}
\vskip -10mm
\caption{Scaling of exotic charmonium hybrids. Expecting mainly ${\cal O}(a_s^2, a_t^2)$ errors, we plot our results against $a_s^2$ (at fixed anisotropy).
Within our (small) statistical errors we cannot resolve any significant 
discretisation errors for the lowest lying exotic $1^{-+}$. }
\vskip -8mm
\label{fig:hybrids_vs_as2}
\end{figure}

In Figure \ref{fig:hybrids_vs_as2} we plot our results with fixed renormalised
anisotropy, $\xi=2$, and different spatial lattice spacing, $a_s$.
It is apparent that there are only small discretisation errors and 
we estimate 4.294(71)(200) GeV for the energy of the lowest lying exotic 
charmonium  excitation, $1^{-+}$, in good agreement with other relativistic 
\cite{milc} and non-relativistic \cite{nrqcd_hybrid} lattice calculations.
The second error is an estimate for quenching errors
which presents the biggest uncertainty in our calculation.
We also expect many non-exotic hybrids in this energy region, 
but they will mix with conventional states, both experimentally 
and on the lattice. The question whether stable hybrids may ultimately 
be found just below the threshold into $D_1 \bar D$-decay (4.28 GeV) 
will have to be decided in a full dynamical simulation.

We also find two other exotics, $0^{+-}$ and $2^{+-}$, 300-500 Mev above 
the $1^{-+}$, as predicted from flux tube models \cite{flux_tube}.
Our attempts to measure the $0^{--}$ exotic were unsuccessful -
presumably this excitation is too high to be resolved on our lattices
\cite{sum_rules}.

\begin{figure}[t]
\hbox{\epsfxsize = 70mm  \epsfysize = 60mm \hskip -1mm \epsffile{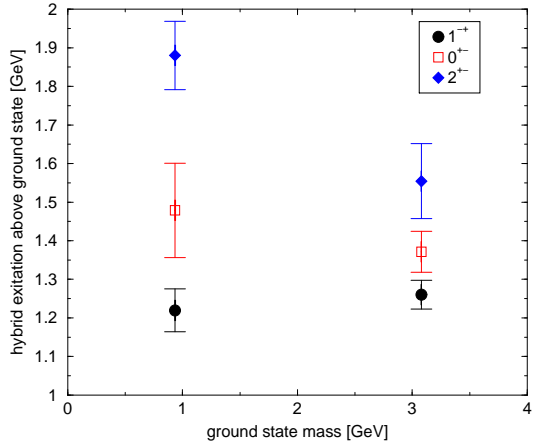}}
\vskip -8mm
\caption{Hybrid excitations above the ground state. For the lowest lying exotic one can observe only a mild mass dependence.}
\vskip -5mm
\label{fig:hybrid_vs_mass}
\end{figure}

In Figure \ref{fig:hybrid_vs_mass} we observe a mild mass dependence
of the exotic spectrum, but the characteristic level ordering $1^{-+} < 0^{+-} < 2^{+-}$ is the same for the whole mass range between $\phi(s\bar s)$ and $J/\Psi(c \bar c)$.

NRQCD simulations have observed a similar mass independence of 
the hybrid levels at even higher masses between charmonium and bottomonium
\cite{nrqcd_hybrid}.
We take this as indication that in the quenched theory the hybrid levels
are almost completely determined by the gluon dynamics (and not the quarks).

\begin{figure}[t]
\hbox{\epsfxsize = 64mm  \epsfysize = 50mm \hskip -1mm \epsffile{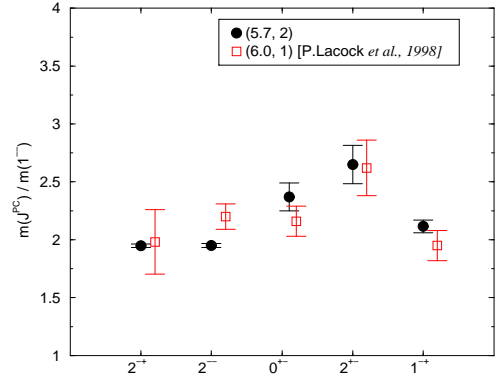}}
\vskip -7mm
\caption{Comparison of the high excitations in $s\bar s$ with previous work
on isotropic lattice \cite{cm_hybrid}. As in that study we plot our results 
normalised with respect to the $1^{--}$.}
\vskip -6mm
\label{fig:strange_compare}
\end{figure}

In Figure \ref{fig:strange_compare} we also compare our results with
strange quark mass to a simulation from an isotropic lattice \cite{cm_hybrid}.
In contrast to those early results we are now able to extract a clear
level ordering for the lowest lying exotics. We should keep in mind that 
our predictions for the level splittings are all within the quenched approximation and we should expect deviations of $10-20 \%$ from the real world.  

\section{RELATIVISTIC BOTTOMONIUM}
\begin{figure}
\vskip -2mm
\hbox{\epsfxsize = 65mm  \epsfysize = 58mm \hskip -1mm \epsffile{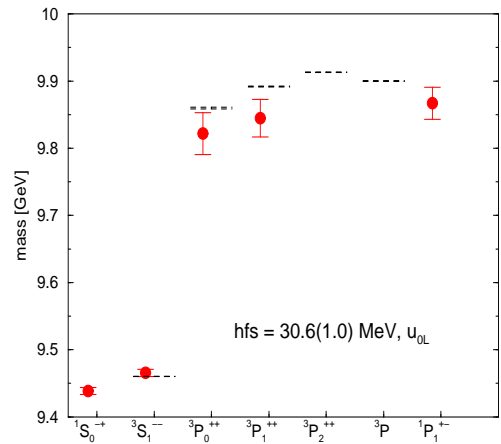}}
\vskip -7mm
\caption{Bottomonium spectrum at $(\beta,\xi) = (6.1, 4)$. The lattice spacing is set from $r_0$.}
\vskip -5mm
\label{fig:bottom_spect}
\end{figure}

Using the same formalism as described in Section 2, we are also able
to perform a relativistic calculation of bottomonium where $m_b \approx 5$ GeV. To this end we simply adjust the anisotropy to have $m_b a_t < 1$.
For our initial study we chose $(\beta,\xi)=(6.1, 4)$ which corresponds 
to $a_s \approx 0.1$ fm and $a_t \approx 0.025$ fm. The low lying spectrum
of bottomonium is shown in Figure \ref{fig:bottom_spect}. 

We are very encouraged by those results and note the good agreement
of the hyperfine splitting ($\approx 30$ MeV) with NRQCD simulations 
at this spatial lattice spacing \cite{nrqcd_omv4,nrqcd_omv6,nrqcd_spitz}. 
But in contrast to those results in our approach we can control scaling violations by taking the continuum limit. Work is under way to study the 
bottomonium spectrum on still finer lattices.

In conclusion, we have demonstrated the efficiency of quenched anisotropic 
lattice calculations.
A fine temporal resolution is paramount to treat heavy quarks in
a fully relativistic framework. Here we also took advantage of a 
small temporal lattice spacing to extract high excitations in the 
spectrum for the whole range of quark masses from strange to bottom.



\end{document}